# Thermal Conductivity of MgO at High Pressures

Ronald E. Cohen
*Carnegie Institution of Washington and Center for High Pressure Research, Geophysical Laboratory,
Washington, D.C. 20015*

The first non-empirical prediction of lattice thermal conductivity of MgO has been determined using molecular dynamics (MD), a non-empirical ionic model (the Variationally Induced Breathing (VIB) model), and Green-Kubo theory. The computation is first-principles is the sense that no parameters are fit to experiment. Results are presented at low pressure as a function of temperature, and for 2500K for pressures to 290 GPa. We find an unexpectedly small pressure effect at small compressions, perhaps due to saturation of thermal conductivity at the high temperatures due to the small mean free path. At higher pressure expected behavior is found.
[thermal conductivity, pressure, molecular dynamics, MgO, computer simulation]

## 1. Introduction

Thermal conductivity is an important parameter for understanding the dynamics and evolution of the Earth, in that it governs heat flow into the mantle from the Earth's core. Pressure effects on thermal conductivity are extremely difficult to measure, and there are only a couple measurements using conventional methods to about 5 GPa[1,2]. The only higher pressure estimates of thermal conductivity were obtained by modeling heat transfer in laser heated diamond anvil experiments on Fe embedded in MgO and $Al_2O_3$, and a zero pressure average value of 4.3% $GPa^{-1}$ was obtained [3] from data at 58 and 125 GPa. There have also been a number of theoretical estimates of pressure effects [4] using Debye theory [5], which gives

$$k = \frac{1}{3} C \bar{v} l \qquad (1)$$

where C is the specific heat per volume, $\bar{v}$ is the average phonon velocity, and $l$ is the mean free path.

## 2. Method

Here, thermal conductivity is computed using molecular dynamics and a very well-tested non-empirical ionic model. The variationally induced breathing model (VIB) is an *ab initio* Gordon-Kim model [6] based on the density functional theory [7]. There is no input from experiment; rather the crystal charge density is modeled as overlapping $O^{2-}$ and $Mg^{2+}$ ions and the total energy is computed using the local density approximation (LDA) [8]. Since $O^{2-}$ is not stable in the free state, the ion is stabilized by a "Watson-sphere," a 2+ charged sphere included in the atomic quantum computations [9]. In the VIB model, the radii of the Watson spheres is chosen by minimizing the total energy as a function of atomic positions [10]. We have found that this model does an excellent job for the thermal equation of state of MgO [11], gives melting behavior [12,13] consistent with other potentials [14] (though not with the experimental melting curve [15]), and gives excellent results for vacancy formation and diffusion in MgO [16]. The overlapping ion charge density and band structure from the potential has been shown to agree closely with self-consistent charge densities for MgO[17]. The total energy in the VIB model is the sum of three parts, the self-energies of each ion (we use the Kohn-Sham self-energy of each ion, with the interaction energy with the Watson sphere removed), the Madelung energy for a lattice of point 2+ and 2- ions, and the short range interaction energy which is the sum of corrections to the Madelung energy for overlap (electron-nuclear and electron-electron), the exchange and correlation overlap energy (using Hedin-Lundqvist [8]), and the Thomas-Fermi kinetic energy. The Watson sphere radii are optimized at each time step. Constant NVE molecular dynamics simulations were performed, in which Newton's equation $F = ma$ is integrated forward in time using fifth-order Gear algorithm [18].

The Green-Kubo formalism was used, which relates the thermal conductivity to fluctuations in the energy current

$$k = \frac{V}{k_B T^2} \int_0^\infty dt \langle J(t) J(0) \rangle, \qquad (2)$$

where $V$ is the volume, $T$ is the absolute temperature, $t$ is time, and $J$ is the energy current [19,20]. The brackets represent a statistical average over the system. It is necessary to define the energy current in terms of the microscopic model. This requires saying where is the energy, or the local energy density. The underlying philosophy is that energy is accounted for where it makes most physical sense and is tractable. Therefore the self-energy is positioned at each nucleus, and the interaction energy between two atoms is divided equally between the two atoms. Then the energy current is given by

$$\vec{J} = \sum_a S_a \vec{v}_a - \frac{1}{2} \sum_{b \neq a} (\vec{F}_{ab} \cdot \vec{v})\vec{r}_{ab} + \sum_{b \neq a} \left(\frac{\partial \phi_{ab}}{\partial P_{wa}}\right)\left(\frac{\partial P_{wa}}{\partial t}\right)\vec{r}_{ab}, \qquad (3)$$

where $S_a$ is the self-energy of ion $a$, $v_a$ is the velocity of ion $a$, $F_{ab}$ is the contribution of the pairwise force between atoms $a$ and $b$, $\phi_{ab}$ is the pairwise short-range potential between atoms $a$ and $b$, $P_{wa}$ is the Watson sphere potential for atom $a$, and $r$ is the vector between $a$ and $b$. The second term is the normal contribution one obtains in for a rigid ion model, and the first and third terms are due to the many-body nature of the VIB model. The first term is the energy transport from the self-energy of the ions, and the third-term is from the effect of changes of Watson sphere potential on the pair potentials.

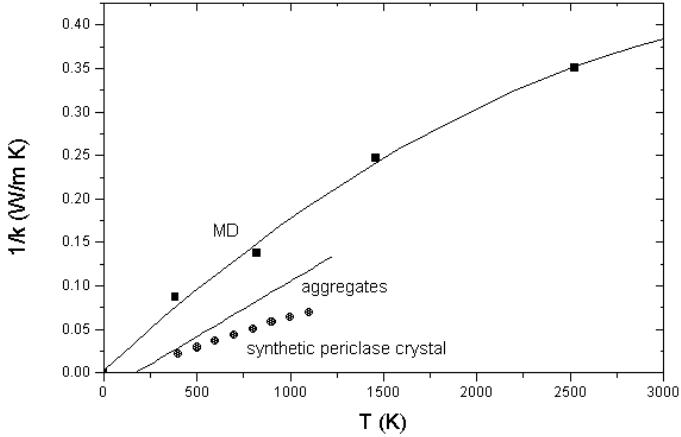

Fig. 1. Inverse thermal conductivity computed using MD compared with experiments on single crystals and aggregates.

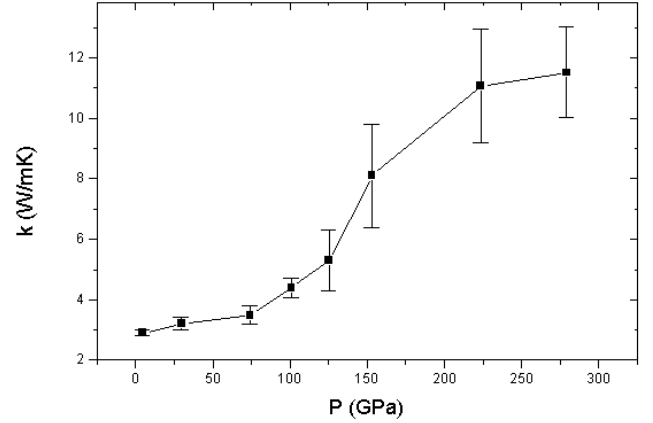

Fig. 2, Predicted thermal conductivity versus pressure at 2500K for MgO using the VIB model and molecular dynamics.

$J(t)$ was obtained from the MD simulations and the autocorrelation function was computed using the Fourier transform method [21] and the thermal conductivity computed. Errors bars were obtained from the statistics of the fluctuations. The system size was 64 atoms in periodic boundary conditions. A limited number of tests indicated that the $k$ was converged at high temperatures for this system size. Most of the runs were done at an energy corresponding to an average temperature of 2500K.

### 3. Results

First, the behavior near zero pressure is considered with respect to temperature. At high temperatures, experimental data can usually be reduced as

$$1/k = a + b\,T, \quad (4)$$

where $a$ is a sample dependant constant primarily extrinsic in nature, depending on defects, impurities, and grain size, and $b$ is related to anharmonicity, or phonon-phonon scattering that leads to a finite thermal conductivity (a purely harmonic crystal would have infinite thermal conductivity). Actually, this simple relationship should be obeyed at constant volume, though much experimental data does fit this form at constant pressure as well. A set of simulations at a constant density of 3.35 g/cm$^3$ were performed at different temperatures (Fig. 1). With a linear fit, there is a small offset ($a \neq 0$ in the eq. 4) and $1/k=0.0419+1.26\times10^{-4}$ T is obtained. However, $a$ should be zero for the perfect crystal. A quadratic fit goes through zero, however, and gives $1/k= 1.97\times10^{-4}$ T$-2.34\times10^{-8}$ T$^2$ (see fig. 1).

The high temperature saturation will be discussed further below. The aggregate data, interestingly, have the same slope as the linear fit to the MD results, with $1/k=-0.0215+1.27\times10^{-4}$ T [22,23] Single crystal experiments [24,23] give $1/k=-0.00531 +6.944\times10^{-5}$ T.

The main goal of this work is to understand pressure effects on thermal conductivity. Fig. 2 shows the computed lattice thermal conductivity versus pressure at 2500K. Only a very moderate increase is found to 130 GPa, pressures corresponding to the base of the mantle. Katsura quotes a relative increase of 3.8% GPa$^{-1}$ over the 5 GPa range of his experiment [1], but re-analyzing his data, I obtain 5±3% 2σ at 373 K and 6±3% 2σ at 1473K. The MD results give a much smaller initial slope of 0.3% GPa$^{-1}$ at 2500 K (or $0.5/K_0$). Manga and Jeanloz analyzed thermal conductivity in terms of power law and linear models, and obtained initial slopes of $(7\pm1)/K_0$ for the linear model, and $(3.0\pm0.7)/K_0$ for the power law model. An initial slope of $7/K_0$ is expected in the Debye model considered by Roufosse and Jeanloz [25], where the thermal conductivity is given by eq. 1, and the mean free path $l$ is assumed to be given by

$$l = \frac{d}{\alpha\gamma T} \quad (5)$$

where $d$ is the interatomic distance, $\alpha$ is the thermal expansivity, and $\gamma$ is the Gruneisen parameter. Typical values then give $(\partial \ln k/\partial \ln \rho)=7$. For comparison in percent change, the linear model gives 5±1% 2σ GPa$^{-1}$, and the power law model 2±1% 2σ GPa$^{-1}$.

These results are analyzed further in fig. 3 where the ln k is plotted versus ln ρ and compared with various models. Manga and Jeanloz considered a power law $k/k_0=(\rho/\rho_0)^7$ and a linear model $(k-k_0)/k_0=7(\rho-\rho_0)/\rho_0$, both consistent with their analysis of the data[3]. Although at high compressions the thermal conductivity begins to behave as expected in the power law model, at low pressures it does not. The linear model also predicts too large an increase at low compressions compared with the MD results. At low pressures a simple proportionality is more consistent with the MD results.

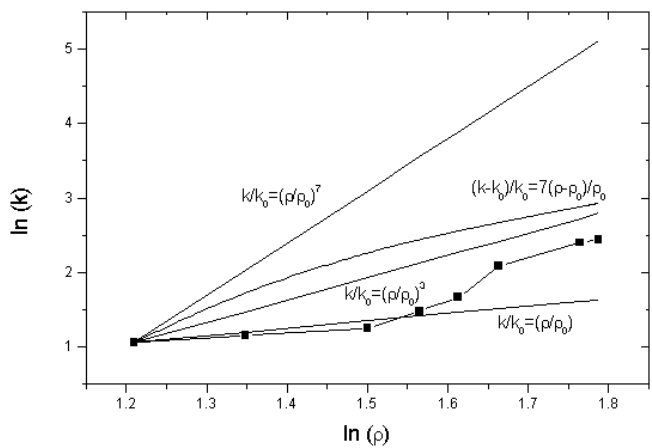

Fig. 3. Predicted thermal conductivity compared with various models.

Considering the difficulty and preliminary nature of the simulations, the difficulties of the experiments, and sparsity of data, it is impossible to draw firm conclusions, but since thermal conductivity at the base of the mantle is an important geophysical parameter, further experimental and theoretical studies are justified. The present results suggest that anharmonicity and phonon-phonon scattering may behave quite differently than has been assumed with compression. The present results show a much smaller initial pressure effect than expected from theory and limited experimental data. Perhaps the low initial slope is due to the high MD temperatures, over 1000K higher than previous low pressure data. One possibility is that the mean free path becomes so small at these temperatures that the thermal conductivity reaches the minimum thermal conductivity where the Debye theory breaks down, and saturates. In fact, using eq. 5 and values from [11] $l$=5 at 2500K, compared with about 75 at 300K, and eq. 5 may underestimate $l$ at very high temperatures. At higher pressures, the mean free path rises due to the decrease in thermal expansivity $\alpha$. The saturation of the thermal conductivity would be consistent with a small pressure dependence at low pressures. The low pressure data (Fig. 1) does indeed seem to indicate saturation at high $T$, and as discussed above, a linear fit gives an unexpected residual term, further bolstering the quadratic behavior. Simulations at lower temperatures as a function of compression would clarify this.

### Acknowledgments

This work is supported by NSF EAR-9418934. M. Kluge gave technical assistance. Much thanks to J. Feldman, I. Inbar, J.M. Brown, and R. Jeanloz for helpful discussions.